# Outstanding figure of merit at high temperature for DFT-based predicted double perovskite oxides, Ba$_2$GaXO$_6$ (X = V, Nb, Ta)

S. S. Saif, M. M. Hossain, M. A. Ali*

*Advanced Computational Materials Research Laboratory, Department of Physics, Chittagong University of Engineering and Technology (CUET), Chattogram 4349, Bangladesh*

**Abstract:**

Thermoelectric materials with a high figure of merit (*ZT*) are highly demanded for a sustainable solution to the energy crisis. In this study, we have predicted three new double perovskite oxides (DPOs), Ba$_2$GaXO$_6$ (X = V, Nb, Ta), with high *ZT* values using density functional theory (DFT) calculations and investigated their structural, electronic, thermoelectric, and mechanical properties. The structural stability was confirmed through the energy-volume curve, octahedral factor, Goldschmidt's tolerance factor, new tolerance factor, formation energy, phonon dispersion spectra, and ab initio molecular dynamics (AIMD) simulations, which indicated the feasibility of synthesizing the predicted compounds. The electronic properties, such as electronic band structure, density of states (DOS), and charge density mapping, are used to disclose the conductive nature, chemical bonding within these compounds, which exhibit direct band gaps of 0.924, 2.354, and 3.279 eV for Ba$_2$GaVO$_6$, Ba$_2$GaNbO$_6$, and Ba$_2$GaTaO$_6$, respectively, as calculated using the TB−mBJ potential. Elastic stiffness constants $C_{11}$, $C_{12}$, and $C_{44}$ analysis confirms mechanical stability and ductile behavior, consistent with the ionic nature of bonding. The thermoelectric performance of the new DPOs, Ba$_2$GaXO$_6$ (X = V, Nb, Ta), was assessed using the BoltzTrap2 code, which yielded outstanding *ZT* values of 2.36, 1.78, and 1.91 at 1500 K for Ba$_2$GaVO$_6$, Ba$_2$GaNbO$_6$, and Ba$_2$GaTaO$_6$, respectively, indicating their potential for waste heat management. The high ZT values are attributed to an ultra-low lattice thermal conductivity, which is due to the strong scattering of acoustic and optical phonon modes. The changes in thermoelectric parameters with temperature were analyzed and explained. As the outcome of this study, the Ba$_2$GaXO$_6$ (X = V, Nb, Ta) perovskites are identified as a promising thermoelectric material, providing a sustainable solution to the current energy crisis.

Corresponding Author: ashrafphy31@cuet.ac.bd

1. **Introduction**

Double perovskite oxides (DPOs) are regarded as a prominent class of materials due to their wide range of compositional and formational possibilities as well as their chemical flexibility, which makes them well-suited to address global energy challenges and reduce dependence on fossil fuels [1, 2]. These materials hold significant aptitude for next-generation photovoltaic devices with enhanced conversion efficiencies and for thermoelectric applications that can convert industrial waste heat into electrical power [3, 4]. The DPOs have the general formula of $A_2BB'O_6$, where *A* is a rare-earth or alkaline-earth metal, *B* and *B'* are transition metals, and *O* represents oxygen [5]. When considering the chemical versatility of potential $A_2BB'O_6$ structures, a wide range of double perovskite oxides can be obtained by employing different cations from the periodic table [6]. Charge equilibrium is ensured by selecting elements that satisfy the condition *$2Q_A + Q_B + Q_{B'} = 12$* [7]. Double Perovskites were first proposed by John B. Goodenough in the 1950s, when he investigated materials that could have a perovskite-like structure with more than one metal ion on the *B*-site. This concept was rooted in the basic perovskite structure, first discovered by Gustav Rose in 1839 in the Ural Mountains and named in honor of Lev Perovski, a Russian mineralogist. The idea of doubling the metal ions at the *B*-sites was novel [8, 9]. The first successful synthesis of a double perovskite oxide occurred in the 1960s. Early studies focused on materials such as $Sr_2FeMoO_6$ and $Sr_2FeReO_6$, where two different transition metal ions (*Fe* and *Mo*, or *Fe* and *Re*) were substituted for each other in the *B*-site of the perovskite structure. These materials demonstrated intriguing electronic and magnetic properties, marking a significant milestone in the study of complex oxides [10].

In recent years, socioeconomic development and rising living standards have brought issues of energy supply and environmental conservation to the forefront of global concerns [11]. The gradual depletion of conventional energy resources, along with the environmental damage caused by their use, poses significant challenges [12]. The quest for sustainable, cost-effective, and efficient energy technologies to mitigate climate change is now a global priority [13]. It is noteworthy that approximately 60% of the energy extracted from conventional energy sources has been dissipated as waste heat [14]. Photovoltaic (*PV*) and thermoelectric (*TE*) technologies offer promising alternative approaches, *PV* system converts solar radiation directly into electricity [15], while *TE* devices can harvest industrial waste heat and vehicle exhaust for power generation [16]. The primary challenge in this field is to significantly improve the energy conversion efficiencies of *PV* and *TE* systems beyond those of traditional technologies. Achieving this advancement depends critically on the discovery and development of novel materials. The performance of a thermoelectric material is characterized by its dimensionless figure of merit [17], $ZT = S^2\sigma T/k_{tot}$. An efficient thermoelectric material requires a high electrical conductivity ($\sigma$), a large Seebeck coefficient (*S*) to maximize heat-to-electricity conversion efficiency, and low thermal conductivity ($k_{tot}$) [14]. The goal is to identify or design materials that meet these combined criteria, satisfying the essential requirements of both optoelectronic and thermoelectric applications.

Numerous $A_2BB'O_6$ type double perovskite oxides have been described in the literature [18]. Their unique structural and electronic properties continue to make them a subject of intensive research today. Rahman *et al.* [1] investigated $Ca_2ZrTiO_6$ and found it to be a non-magnetic semiconductor material with a direct bandgap of 2.3 eV. They calculated a maximum figure of merit (*ZT*) of 4.4 at a temperature of 550 K and concluded that $Ca_2ZrTiO_6$ double perovskite oxide is a highly promising candidate for thermoelectric applications. Dixit *et al.* [19] evaluated the thermoelectric performance of $Ba_2InNbO_6$, reporting that this material achieved a *ZT* value of 0.7 at 1200 K. Khandy *et al.* [20] analyzed $Sr_2HoNbO_6$ theoretically and found it to be a semiconductor with an energy gap of about 3.6 eV; they calculated a maximum figure of merit (*ZT*) of 0.97 at 300K for this material. Ishfaq *et al.* [21] evaluated the thermoelectric performance of $Ba_2CeSnO_6$ and $Ba_2CePtO_6$, reporting that these materials achieve a maximum *ZT* value of 0.66 and 0.74 at 800 K, respectively. They also found that $Ba_2CeSnO_6$ and $Ba_2CePtO_6$ exhibit large Seebeck coefficients of $S \approx 170$ and 221 $\mu$V/K, respectively, highlighting their promise for thermoelectric energy conversion. Haid *et al.* [22] studied $Sr_2CrTaO_6$, identifying it as a half-metallic ferromagnet in the ground state, and reported a *ZT* of 0.6 at 800 K, further highlighting the spintronic potential of double perovskite oxides. Al−Qaisi *et al.* [23] predicted that $Ba_2YBiO_6$ is an indirect p-type semiconductor with a figure of merit (*ZT*) of 0.78 at 600 K, indicating its suitability for thermoelectric (*TE*) applications. Bellahcene *et al.* [24] analyzed the spin-polarized transport properties of $Sr_2PrRuO_6$ and reported a *ZT* value of 0.90 at 1000 K for spin up, highlighting its promise for efficient thermoelectric applications. Dar *et al.* [25] investigated $Ba_2InTaO_6$ and found it to be a semiconductor with electrons as the primary charge carriers, which exhibits a high-power factor ($S^2\sigma$), highlighting its potential for *TE* applications. Compounds with different *A*-site cations have also shown interesting behavior. Aziz *et al.* [26] studied the double perovskites $X_2NaIO_6$ (X

= Pb, Sr), and found that both $Pb_2NaIO_6$ and $Sr_2NaIO_6$ are semiconductors with direct band gaps of 3.75 eV and 5.48 eV, respectively. $Sr_2NaIO_6$ achieved a higher $ZT$ of 0.7728 at 650 K and a large power factor, $PF \approx 206.3$ $\mu W/mK^2$, suggesting excellent $TE$ performance. Both compounds' cubic structures are advantageous for thermoelectric and optoelectronic devices. Hanif et al. [27] examined $Sr_2LuNbO_6$ and $Sr_2LaNbO_6$, reporting direct band gaps of 3.7 eV and 4.02 eV, respectively, with corresponding $ZT$ values of 0.819 and 0.779 at 750 K. O. Sahnoun et al. [28] noted that $Ba_2FeMoO_6$, a half metallic double perovskite, exhibits a remarkably strong thermoelectric response, with $ZT \approx 0.998$ at 200 K. Aziz et al. [29] investigated $Sr_2CaWO_6$ and $Sr_2MgWO_6$ in their cubic phase, showing that both compounds are semiconductors with direct band gaps of 4.4 and 4.3 eV and also calculated the optical properties and explored the transport properties under varying temperature, reporting maximum $ZT$ values of 0.79 at 650 K for $Sr_2CaWO_6$ and 0.78 at 800 K for $Sr_2MgWO_6$, which highlight their potential for renewable energy applications. Experimental studies have confirmed and extended these predictions. Himanshu et al. [30] used synchrotron X-ray diffraction and UV−Visible reflectance to show that $Ba_2ScTaO_6$ crystallizes in an ordered cubic phase Fm−$\bar{3}$m (No. 225) with a wide band gap of about 4.7 eV. Manoun et al. [31] employed Raman spectroscopy to study $Sr_2MgWO_6$ and observed a tetragonal-to-cubic phase transition at approximately 550°C. Kockelamann et al. [32] performed high-resolution neutron powder diffraction and found that $Ba_2PrIrO_6$ is also cubic Fm−$\bar{3}$m (No. 225) at room temperature, and that it undergoes anti-ferromagnetic ordering below 71 K. Synthesis of double perovskite oxides is comparatively easy via solid-state reaction techniques. Chen et al. [33] synthesized $Ba_2PrRuO_6$ by solid-state reaction and, from powder X-ray diffraction, identified an ordered cubic structure with space group Fm−$\bar{3}$m (No. 225) at room temperature. Chang et al. [34] prepared $La_2CoMnO_6$ via solid-state synthesis and reported an orthorhombic structure, Pnma (No. 62), with a band gap of 1.93 eV. Aziz et al. [35] employing the conventional solid-state route to fabricate $Sr_2TiCoO_6$, determined a monoclinic $P2_1/n$ crystal symmetry and reported a band gap of 2.03 eV. We have mainly been motivated by reporting a high $ZT$ value (4.4 at 550 K) for $Ca_2ZrTiO_6$ [1] and also found $ZT$ values are reasonably good (close to 1.0) for other double perovskites [19−29]; consequently, we aimed to design and discover some new double perovskites with high $ZT$ values. In this journey, we have screened a large number of DPOs, wherein some exhibit very low $ZT$ values, while others are unstable. Finally, we obtained three new compounds with very good $ZT$ values: $Ba_2GaVO_6$ exhibits an outstanding $ZT$ value of 2.36 at 1500 K, while $Ba_2GaNbO_6$ and $Ba_2GaTaO_6$ exhibit ZT values of 1.78 and 1.91, respectively, at the same temperature. Noted that the predicted compounds satisfy the stability criteria and are expected to be synthesizable.

Therefore, this study focuses on the initial screening of the stable double oxide perovskites $Ba_2GaXO_6$ (X = V, Nb, Ta) and investigates their structural, electronic, thermoelectric, and mechanical properties. The combination of non-toxic composition, favorable direct band gaps, and high *ZT* values makes these compounds as highly promising candidates for thermoelectric and device applications. Our goal is to evaluate their potential by characterizing the key physical properties that determine device performance.

## 2. Computational method:

The double perovskite oxide compounds $Ba_2GaXO_6$ (X = V, Nb, Ta) were investigated using first principles density functional theory (DFT) [36, 37] calculations within the full potential linearized augmented plane wave (FP−LAPW) method as implemented in the WIEN2k code [38]. Exchange correlation effects were treated using the Perdew−Burke−Ernzerhof (PBE) [39] generalized gradient approximation [40], and the modified Becke−Johnson (TB−mBJ) [41] potential was also applied to improve the accuracy of the electronic structure. In the FP−LAPW calculations, wavefunctions inside the non-overlapping muffin-tin spheres were expanded up to $I_{max}$ = 10, while in the interstitial region, a plane wave cutoff of $R_{MT} \times K_{max}$ = 8, $G_{max}$ = 12, where $R_{MT}$ is the smallest muffin-tin radius, was used to define the basis set. An energy cutoff of −6.0 *Ry* was set to separate core and valence states. Brillouin zone integrations were carried out using a Monkhorst-Pack [42] k-point grid of 12×12×12, without shifting, to achieve self-consistency when the total energy convergence (ec) between iterations was less than $10^{-5}$ Ry and the charge convergence (ce) was 0.0001 e. Density of state (DOS) was computed using a denser non-shifted *k*-point grid of 20 × 20 × 20, and the thermoelectric transport parameters were obtained with a denser 45 × 45 × 45 mesh by means of the BoltzTraP2 code under Boltzmann's theory [43], assuming a constant relaxation time, $\tau$ = $10^{-14}$ s [44] and the rigid band approximation. Phonon dispersion relations were calculated using the CASTEP code with a finite displacement approach to assess the dynamical stability [45]. To further validate the dynamical stability, we performed 10 ps ab initio molecular dynamics (AIMD) simulations using the CASTEP code [45].

## 3. Results and discussion

### 3.1 Structural properties and stability criteria

The double perovskite oxides $Ba_2GaXO_6$ (X = V, Nb, Ta) adopt the ideal cubic perovskite structure, space group Fm−3̄m (No. 225) with a face centered cubic unit cell containing four formula units in the ratio of 2:1:1:6, which is depicted Fig. 1. In this framework the *Ba, Ga, X,* and *O* atoms occupy the 8c, 4b, 4a, and 24e Wyckoff sites at fractional coordinates (0.25, 0.75, 0.75), (0, 0, 0.5), (0, 0, 0), and (0.230779, 0, 0.5), respectively. Each $Ba^{2+}$ cation is coordinated by twelve $O^{2-}$ anions, forming $BaO_{12}$ cuboctahedra that share corners with neighboring $BaO_{12}$ units and faces with $GaO_6$ and $XO_6$ octahedra. The structural parameters and atomic positions were optimized using density functional theory with the PBE−GGA functional by minimizing the total energy $E_{tot}(V)$ as a function of the primitive cell volume. The equilibrium lattice constant, bulk modulus, $B_0$, first derivative of bulk modulus, $B_0'$, and volume, $V_0$, were extracted by fitting the calculated $E_{tot}(V)$ data to the Birch−Murnaghan equation of state [46]. The Birch−Murnaghan equation of state (EOS) is given by:

$$E_{tot}(V) = E_o(V) + \frac{B_0 V}{B_0'(B_0' - 1)}\left[B\left(1 - \frac{V_0}{V}\right) + \left(\frac{V_0}{V}\right)^{B_0'} - 1\right] \quad (1)$$

Where $B_0$ is the bulk modulus, and $B_0'$ its derivative, $E_0$ is the minimum energy at equilibrium volume $V_0$.

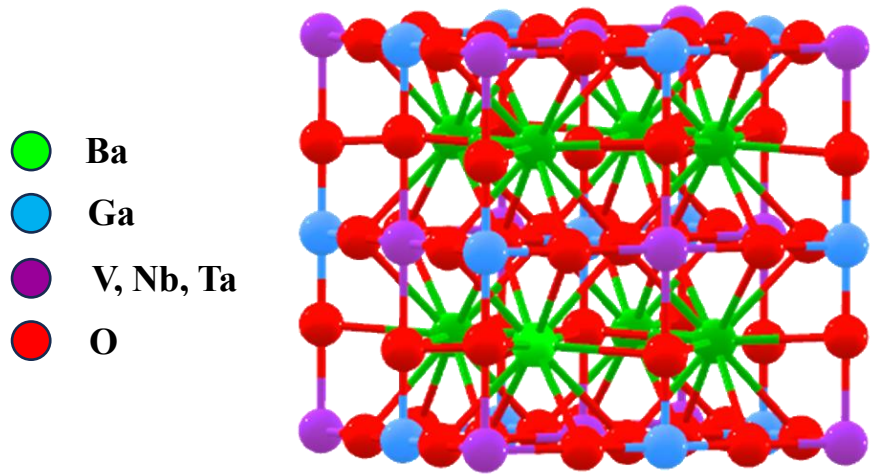

**Fig. 1.** Crystal structure of $Ba_2GaXO_6$ (X = V, Nb, Ta).

Fitting $E_{tot}(V)$ to this form yields the ground state structural parameters listed in Table 1. These equilibrium parameters are illustrated by the optimized $E_{tot}(V)$ optimization curves (Birch−Murnaghan fits) shown in Fig. 2. When computing equilibrium properties of a crystal, the ground-state total energy is found at the minimum of the energy-volume curve. For a perfect cubic perovskite, the lattice constant at this minimum energy (optimized) volume gives the equilibrium lattice parameter of the structure [47].

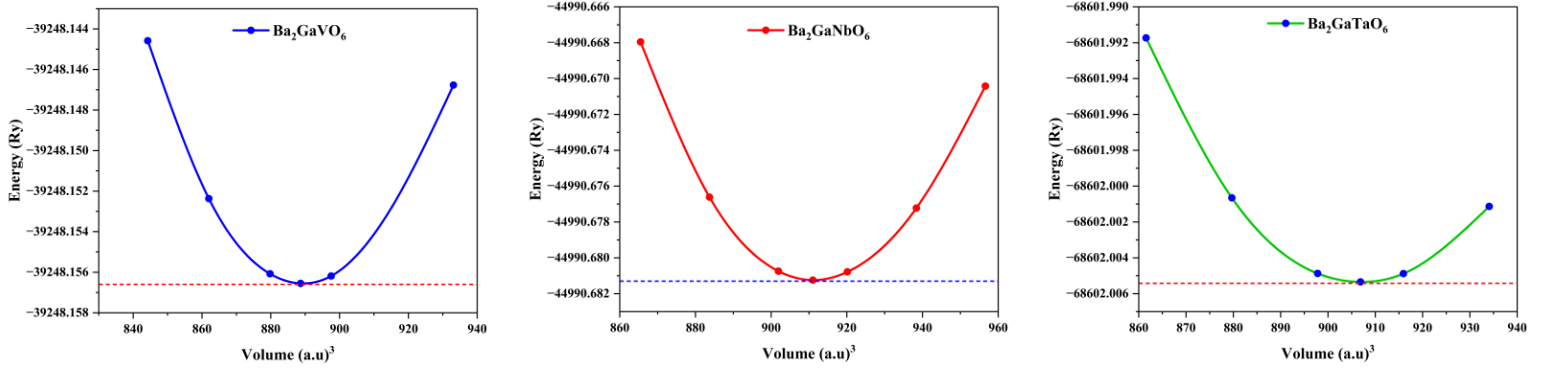

**Fig. 2.** Calculated total energy as a function of volume $E(V)$ for $Ba_2GaXO_6$ (X = V, Nb, Ta) compounds, optimized structure using PBE−GGA.

Assessing the stability of a crystal structure is crucial for potential applications. A traditional measure for perovskite stability is Goldschmidt's tolerance factor ($T_F$), which provides a simple size-based criterion [48]. Marina et al. later introduced an octahedral factor ($\mu$) to refine this criterion [49]. By using $T_F$ and $\mu$ together, one can correctly predict perovskite stability in about 80% of cases [50]. In practice, $T_F$ and $\mu$ are defined in terms of the ionic radii of the constituent atoms according to the conventional Goldschmidt-type expressions as follows:

$$T_F = \frac{R_A + R_O}{\sqrt{2}\left(\frac{R_{B'} + R_{B''}}{2} + R_O\right)} \qquad (2)$$

$$\mu = \frac{R_{B'} + R_{B''}}{2 R_O} \qquad (3)$$

Here, Shannon's ionic radius for the elements is expressed by R, $R_A$ = $Ba^{2+}$ with coordination number 12, $R_{B'}$ = $Ga^{3+}$, $R_{B''}$ = $V^{5+}$/ $Nb^{5+}$/ $Ta^{5+}$, and $R_O$ =$O^{2-}$ with coordination number 6 [51]. Goldschmidt's tolerance factor works well for simple perovskites but is less reliable for more complex double perovskite structures, where additional factors, such as detailed electronic configuration and local coordination, become important. To improve stability predictions for double perovskites, we have applied a modified tolerance factor that incorporates the specific structural and electronic characteristics of Ba$_2$GaXO$_6$. This extended criterion extends beyond simple ionic size ratios by incorporating additional parameters, thereby providing a better match to experimental stability observations. Bartel et al. [52] introduced a new tolerance factor, $\tau$, which achieves roughly 92% accuracy in predicting perovskite stability. $\tau$ is defined as follows:

$$\tau = \frac{2R_O}{R_{B'} + R_{B''}} - n_A \left[ n_A - \left( \frac{2R_A}{R_{B'} + R_{B''}} \right) \bigg/ \ln \left( \frac{2R_A}{R_{B'} + R_{B''}} \right) \right] \quad (4)$$

Where $n_A$ is the oxidation state of the A-site, here $n_A = +2$ for Ba [53]. Stable perovskites typically satisfy $0.81 \leq T_F \leq 1.11$ [54], $0.41 \leq \mu \leq 0.90$ [55], and $\tau < 4.18$ [56], which is taken as an indicator of stability for double perovskite compounds. The calculated values of $T_F$, $\mu$, and $\tau$ for Ba$_2$GaVO$_6$, Ba$_2$GaNbO$_6$, and Ba$_2$GaTaO$_6$ are listed in Table 1. For Ba$_2$GaXO$_6$ (X = V, Nb, Ta) compounds, all of the calculated values of $T_F$, $\mu$, and $\tau$ fall within the empirically established stability ranges. Therefore, these materials are expected to form stable perovskite structures.

**Table 1**: Calculated structural parameters of Ba$_2$GaXO$_6$ (X = V, Nb, Ta) compounds.

| Parameter | Ba$_2$GaVO$_6$ | Ba$_2$GaNbO$_6$ | Ba$_2$GaTaO$_6$ |
|---|---|---|---|
| Lattice constant, $a = b = c$ (Å) | 8.0761 | 8.1433 | 8.1308 |
| Volume, $V_0$ (Å$^3$) | 526.751 | 540.009 | 537.526 |
| Bulk Modulus, $B$ (GPa) | 164.142 | 151.509 | 158.934 |
| First derivative of bulk modulus, $B'$ (GPa) | 5.2424 | 4.7092 | 4.7137 |
| Ground-state total energy, $E_0$ (Ry) | -39248.16 | -44990.68 | -68602.005 |
| Goldsmith's tolerance factor, $T_F$ | 1.075 | 1.048 | 1.048 |
| octahedral factor, $\mu$ | 0.414 | 0.450 | 0.450 |
| New tolerance factor, $\tau$ | 3.85 | 3.67 | 3.67 |
| Formation energy, $\Delta H_f$ (eV/atom) | −3.63 | −3.86 | −3.66 |

We evaluated structural stability by calculating the formation energy using the following equation:

$$\Delta H_f = \frac{\left[ E_{Ba_2 GaXO_6} - (8E_{Ba} + 4E_{Ga} + 4E_X + 24E_O) \right]}{40} \quad (5)$$

In this expression $E_{Ba_2GaXO_6}$ denotes the total energy of the compounds, while $E_{Ba}$, $E_{Ga}$, $E_X$, and $E_O$ represent the total energies of *Ba*, *Ga*, *X* (*V, Nb, Ta*), and *O* atoms. As reported in Table 1, the formation energies are negative for all studied compounds, indicating their structural stability [56].

The phonon dispersion relations of a crystal provide key insights into its dynamical stability and thermal properties. In particular, the phonon spectrum reveals whether any vibrational modes have imaginary (negative) frequencies, which would signal a tendency toward structural distortion or phase transition. As depicted in Fig. 3, the phonon dispersion spectra of $Ba_2GaXO_6$ (X = V, Nb, Ta) were computed along the Brillouin zone's high symmetry paths. In a dynamically stable crystal, all phonon branches remain at positive frequencies throughout the Brillouin zone [57]. Conversely, any branches dipping below zero indicate soft modes and an unstable lattice. For each compound, the conventional cubic cell contains four formula units along with 40 atoms in total, yielding 3N = 120 phonon modes [57]. Of the 120 phonon modes, 3 of them are acoustic modes, which reach zero frequency at the $\Gamma$ point as required by translational invariance, and the remaining 117 are optical modes. Acoustic phonons correspond to in-phase vibrations of the lattice, while optical phonons involve out-of-phase motion of atoms against each other. Consistent with general trends, the highest optical branches in $Ba_2GaXO_6$ (X = V, Nb, Ta) occur at the zone center ($\Gamma$ point). Importantly, the calculated phonon spectra depicted in Fig. 3 show no imaginary frequencies for any of the three compounds. All phonon branches remain above zero energy across the entire Brillouin zone, indicating the absence of soft phonon modes. This confirms that $Ba_2GaVO_6$, $Ba_2GaNbO_6$, and $Ba_2GaTaO_6$ are dynamically stable in the cubic perovskite structure.

In each case, there are three acoustic branches emerging from the $\Gamma$ point that extend up to only a few terahertz (THz) before the first optical bands appear. For $Ba_2GaVO_6$, the acoustic modes reach a maximum of 4.5 THz, whereas the heavier compounds, $Ba_2GaNbO_6$ and $Ba_2GaTaO_6$, have a slightly lower acoustic cutoff of 4.2 THz and 4.0 THz. The acoustic branches are relatively flat, reflecting the large masses and modest bond stiffness. Above the acoustic cutoff, multiple optical phonon groups appear. A cluster of lower-frequency optical modes spans roughly 5-8 THz, followed by higher-frequency bands that extend to the top of the spectrum. No large phonon band gap is seen; the acoustic and optical bands are separated by only 1−2 THz. In particular, the optical spectrum is dense, with many modes lying near the acoustic cutoff.

The ultra-low lattice thermal conductivities of these compounds follow directly from the phonon spectra. All three compounds have heavy atoms and a complex unit cell, which suppresses acoustic group velocities and introduces numerous scattering channels. In multi-cation oxides, phonon scattering is greatly enhanced by the large mass and bond strength contrast among the constituents. The numerous low-frequency optical branches further enhance scattering, and the interference between multi-component materials and acoustic and optical modes can significantly dominate thermal resistance. This means that the heat-carrying acoustic phonons readily interact with the broad spectrum of optical phonons, shortening phonon lifetimes and reducing lattice thermal conductivity. $Ba_2GaVO_6$ exhibits the smallest $k_L$ value compared to $Ba_2GaNbO_6$ and $Ba_2GaTaO_6$. The $Ba_2GaVO_6$ compound's optical branches lie especially close to the acoustic band, so that many optical modes (associated with Ga−V−O vibrations) intrude into the low frequency range. Systems with many low-lying optical branches exhibit strong acoustic-optical scattering, which can become the primary limiting factor for heat transport [58]. Thus, $Ba_2GaVO_6$ likely experiences more acoustic-optical mode mixing and broader phonon linewidths than the Nb/Ta analogues. In contrast, $Ba_2GaNbO_6$ and $Ba_2GaTaO_6$ have less optical crowding near the acoustic cutoff, and their acoustic-optical overlap is smaller than that of $Ba_2GaVO_6$, yielding a slightly higher $k_L$. The ultra-low $k_L$ of $Ba_2GaVO_6$ arises from its densely packed low-frequency phonon spectrum and strong anharmonic scattering channels, which severely limit phonon lifetimes and mean free paths.

Each of the examined compounds displays a high entropy value, indicating strong anharmonic behavior and very low lattice thermal conductivity, which enhances their stability at high temperatures [59]. A high entropy ($S$) reflects increased disorder and makes the Gibbs free energy ($G = H − TS$) more negative, especially at higher temperatures [60].

$$S = \left(\frac{\partial G}{\partial T}\right)_P \qquad (6)$$

The calculated data presented in Table 4 reveal entropy values of 233.84 J/K.mol for $Ba_2GaVO_6$, 221.54 J/K.mol for $Ba_2GaNbO_6$, and 225.13 J/K.mol for $Ba_2GaTaO_6$. The relatively higher entropy observed for $Ba_2GaVO_6$ indicates strong anharmonicity, which is strongly correlated with its reduced lattice thermal conductivity and enhanced thermal stability at elevated temperatures.

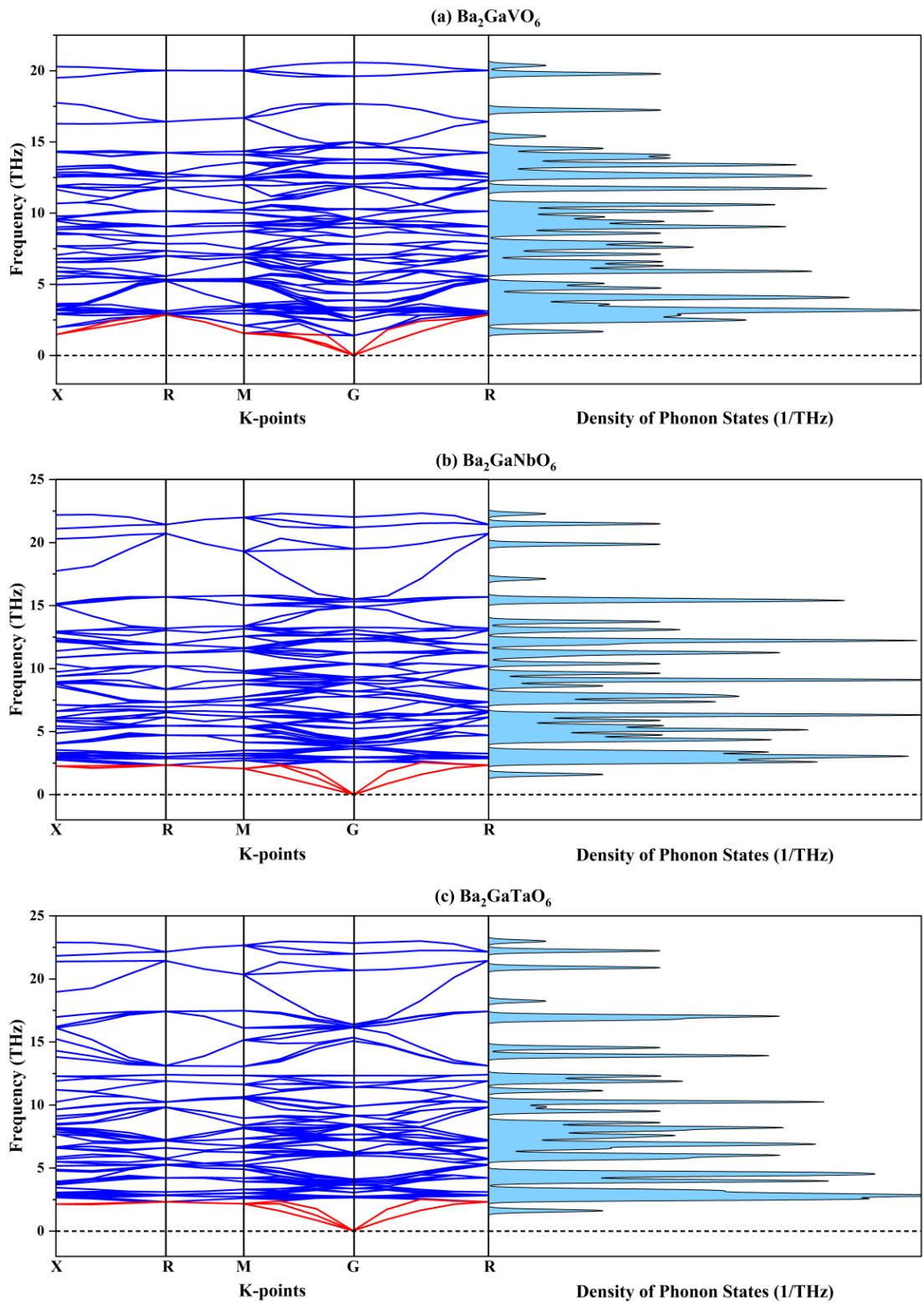

**Fig. 3.** Phonon dispersion spectra (PDS) with the total phonon density of states (TDOS) of $Ba_2GaXO_6$ (X = V, Nb, Ta) compounds at ambient pressure.

To further validate the dynamical stability of the predicted $Ba_2GaXO_6$ (X = V, Nb, Ta) phases, we performed 10 ps, ab initio molecular dynamics (AIMD) simulations, depicted in Fig. 4. The total energy of each compound fluctuated only minimally during the entire simulation. In fact, the maximum energy fluctuations were only 0.0034% for $Ba_2GaVO_6$, 0.00045% for $Ba_2GaNbO_6$, and 0.0084% for $Ba_2GaTaO_6$. These extremely small fluctuations (below 1%) indicate that the structures experienced no significant distortion or decomposition over the simulation period. Such negligible energy oscillations indicate the thermal stability of these compounds. The very low energy drift observed here confirms that all three compounds retain their structural integrity under dynamical conditions [57].

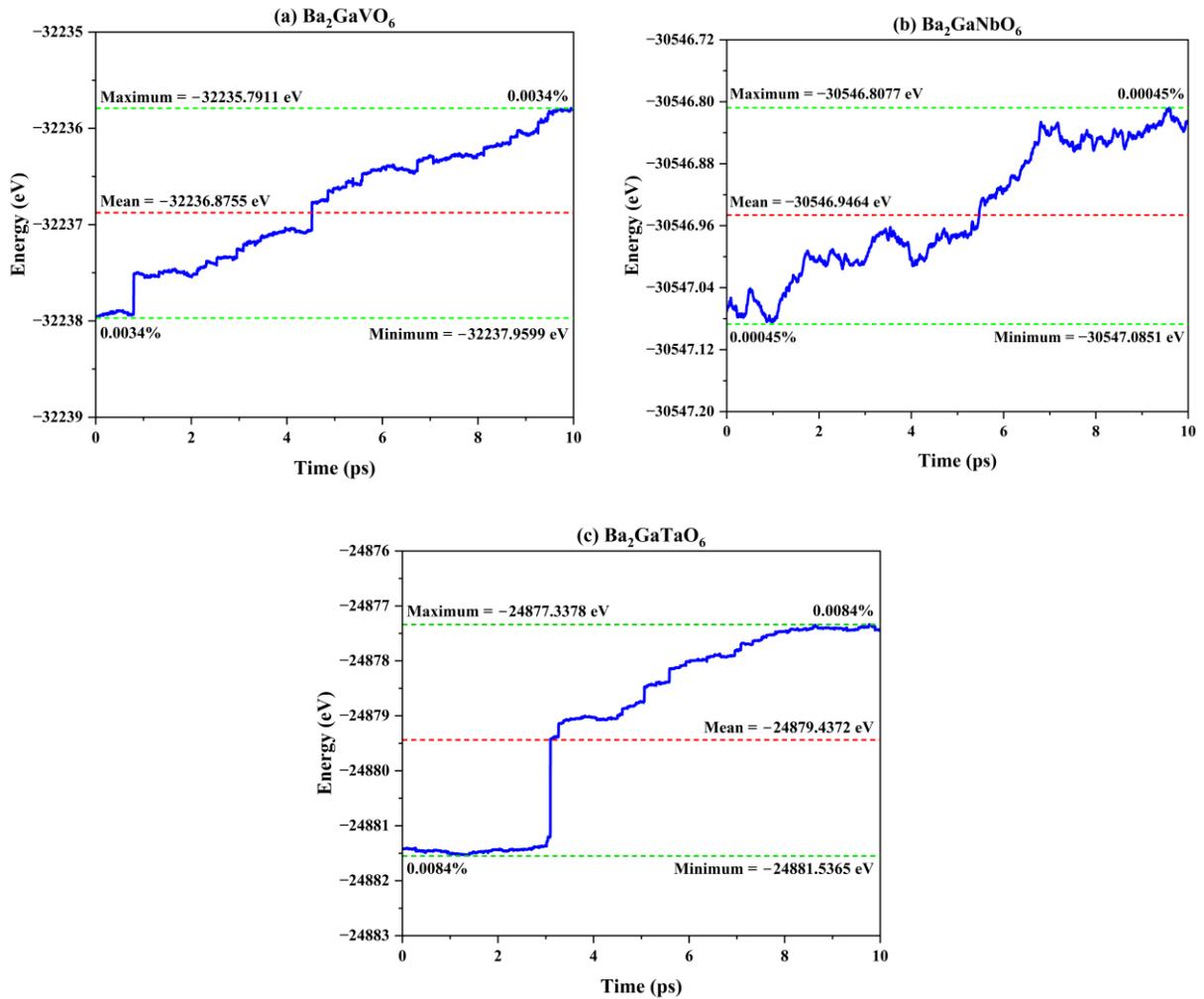

**Fig. 4.** ab initio molecular dynamics (AIMD) of $Ba_2GaXO_6$ (X = V, Nb, Ta) compounds.

Lattice parameters presented above suggest that PBE-GGA optimization predicts a stable cubic Fm−$\overline{3}$m phase for $Ba_2GaXO_6$ (X = V, Nb, Ta). The octahedral factor, $\mu$, tolerance factor, $T_F$, new tolerance factor, $\tau$, formation energy, phonon dispersion curves, and AIMD results are consistent with the double perovskite stability, suggesting these predicted double perovskite oxides are stable and synthesizable.

**3.2. Electronic properties:**

The electronic band structure refers to the spectrum of allowed and forbidden energy states for electrons in a crystalline solid. The highest occupied band is the valence band, and the lowest unoccupied band is the conduction band; the energy difference between their edges determines a material's electronic class (insulator, semiconductor, or metal) [56]. In optoelectronic applications, photons can be absorbed to produce electron-hole pairs only when their energy is equal to or exceeds the band gap ($E_g$) [61]. Photons with energy below $E_g$ pass through as if transparent, while photons just above $E_g$ are efficiently absorbed [62]. The conversion of waste heat into electricity using thermoelectric (*TE*) materials is a highly appealing concept. Achieving efficient heat-to-electricity conversion remains challenging. It requires detailed knowledge of the *TE* material's electronic structure, particularly in the critical band-gap region. This region critically influences the transport properties that govern the heat-to-electricity conversion process. Our calculated band structures for $Ba_2GaXO_6$ (X = V, Nb, Ta) show semiconductor behavior in all cases. The band structure was computed using the PBE-GGA and TB-mBJ exchange-correlation functionals.

The Fermi level ($E_F$) is designated as the reference energy, with its value set to zero. The band structure patterns derived using the TB−mBJ functional are illustrated in Fig. 5, and the calculated results are presented in Table 2. The TB−mBJ band gaps are 0.924 eV, 2.354 eV, 3.279 eV for $Ba_2GaVO_6$, $Ba_2GaNbO_6$, and $Ba_2GaTaO_6$, respectively, which are systematically higher than the PBE−GGA calculated band gaps 0.524 eV, 1.873 eV, 2.420 eV. The PBE−GGA functional typically predicts lower band gaps [63, 64]. It has been reported by A. Boutramine *et al.* that the band gap energy of $Ba_2CePtO_6$ is underpredicted by the PBE−GGA functional by approximately 9% [14], whereas the TB−mBJ functional provides improved band gap predictions. TB−mBJ yields improvements over PBE in all cases [65]. In our results, this manifests as roughly a 0.4 − 0.9 eV increase in the calculated $E_g$. Our calculated band gap values follow the same trend observed for several materials exhibiting large thermoelectric figures of merit (*ZT*). $Cr_2ZrTiO_6$ ($E_g$

= 2.3 eV) has been reported with $ZT$ = 4.4 at 550 K [1], $Sr_2HoNbO_6$ ($ZT$ = 0.97 at 300K) have a bandgap of 3.6 eV [20], $Bi_2O_2Se$ ($E_g$ = 2.16 eV) shows $ZT$ = 3.35 at 800 K [66], $Ga_2O_2$ ($ZT$ = 6.5 at 500 K) has a band gap of 2.77 eV [67], $SrIn_2C_2$ and $BaIn_2C_2$, with band gaps of 0.763 and 0.932 eV, respectively, have reported $ZT$ values of 1.93 and 2.86 at 1000 K [68]. These comparisons suggest that larger band gaps do not necessarily preclude high calculated ZT values. The TB−mBJ band structures (Fig. 5) reveal that for each compound, the valence band maximum (VBM) and conduction band minimum (CBM) are located at an identical $k$-point, indicating the presence of direct ($X$−$X$) band gaps ($E_g$) [20], and the conduction bands are shifted upward. $Ba_2GaVO_6$'s gap opens from 0.524 to 0.924 eV under TB−mBJ, and $Ba_2GaTaO_6$'s gap from 2.420 to 3.279 eV. Thus, replacing $V \rightarrow Nb \rightarrow Ta$ in the lattice progressively widens the band gap, reflecting the heavier cation's influence on the electronic structure.

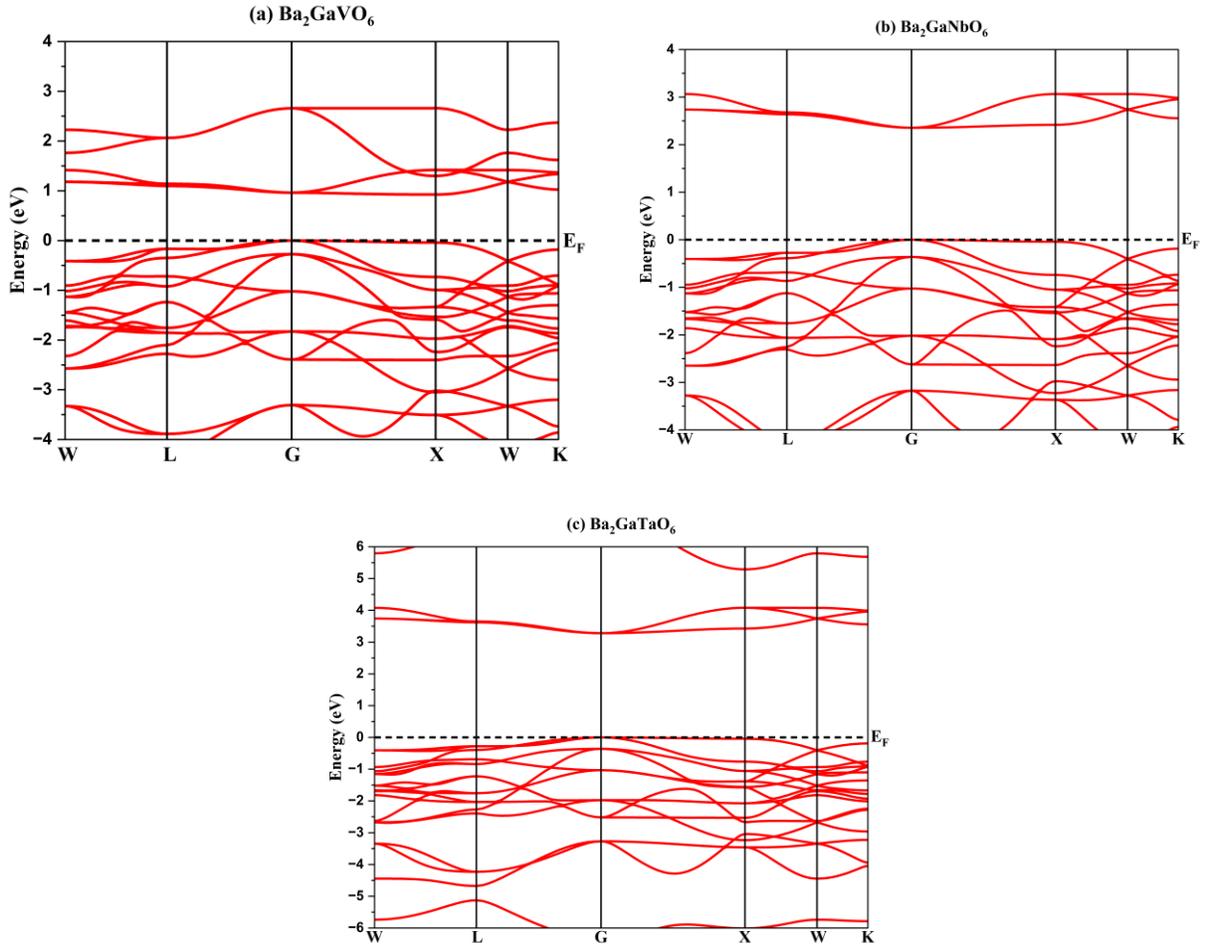

**Fig. 5.** Band structure diagrams of (a) $Ba_2GaVO_6$, (b) $Ba_2GaNbO_6$, and (c) $Ba_2GaTaO_6$.

Ba$_2$GaVO$_6$ exhibits a narrow band gap in the near-infrared range, implying strong absorption of photons across the visible spectrum. The Shockley−Queisser analysis shows that single-junction photovoltaic efficiency peaks around a band gap of ~1.1 eV [69]. While 0.924 eV is slightly lower than this optimum, it remains comparable to silicon (1.1 eV) and would absorb a broad portion of the solar spectrum. Ba$_2$GaVO$_6$'s near-*IR* band gap and direct band behavior make it well-suited as a solar absorber material. By contrast, Ba$_2$GaNbO$_6$ and Ba$_2$GaTaO$_6$ have much wider band gaps in the visible−*UV* range. Such wide gap semiconductors are transparent to visible light and only absorb photons with higher energy. They are also more robust at elevated temperatures, a beneficial trait for high-temperature thermoelectric or waste heat applications [70]. Band gaps of ~2−3 eV fall in the range of well-known photocatalysts, such as TiO$_2$ with $E_g \approx 3.2$ eV [71]. Thus, Ba$_2$GaNbO$_6$ and Ba$_2$GaTaO$_6$ can harness *UV* photons to drive photocatalytic reactions.

**Table 2:** Calculated Band gap values and Carrier Effective Masses of Ba$_2$GaXO$_6$ (X = V, Nb, Ta).

| Compound | TB-mBJ (eV) | PBE-GGA (eV) | $m_e^*(m_0)$ | $m_h^*(m_0)$ | Ref. |
|---|---|---|---|---|---|
| Ba$_2$GaVO$_6$ | 0.924 | 0.524 | 0.79 | 0.25 | This work |
| Ba$_2$GaNbO$_6$ | 2.354 | 1.873 | 0.52 | 0.204 | This work |
| Ba$_2$GaTaO$_6$ | 3.279 | 2.420 | 0.23 | 0.59 | This work |
| Ba$_2$CePtO$_6$ | 1.518 | 1.385 | | | [14] |
| Ba$_2$InNbO$_6$ | 3.634 | 2.965 | | | [20] |
| Ba$_2$AsTaO$_6$ | 3.252 | 2.418 | | | [53] |
| Ba$_2$AsVO$_6$ | 1.676 | 0.528 | | | [53] |

The effective masses of electrons, $m_e^*(m_0)$, and holes, $m_h^*(m_0)$, at the band edges were determined from the slope of the conduction and valence band dispersions. We used the standard relation given below [68]:

$$m^* = \frac{\hbar}{\frac{d^2 \varepsilon(k)}{dk^2}} \qquad (7)$$

The calculated values, including the direct $\Gamma$-point band gaps, are presented in Table 2. Ba$_2$GaVO$_6$ has $m_e^*(m_0) = 0.79$ and $m_h^*(m_0) = 0.25$, Ba$_2$GaNbO$_6$ has $m_e^*(m_0) = 0.52$ and $m_h^*(m_0) = 0.204$, and Ba$_2$GaTaO$_6$ has $m_e^*(m_0) = 0.23$ and $m_h^*(m_0) = 0.59$. Thus, as $X$ goes from V → Nb → Ta, the band gap increases while the calculated electron effective mass decreases sharply, whereas the hole mass shows the opposite trend for *Nb → Ta*. These trends have clear implications for carrier transport. In $Ba_2GaVO_6$, the conduction electrons are relatively heavy, whereas the holes are much lighter. By contrast, $Ba_2GaTaO_6$ has very light electrons but the heaviest holes. $Ba_2GaNbO_6$ lies intermediate in both cases. This indicates that charge transport will be anisotropic in each compound. $Ba_2GaTaO_6$ would be expected to have the highest electron mobility, while $Ba_2GaNbO_6$ should have the highest hole mobility. Conversely, the heavy electrons in $Ba_2GaVO_6$ imply relatively low electron mobility and may conduct holes much more easily. These contrasting effective masses are important for device performance. Notably, lighter bands (high curvature) yield smaller *m\** and faster carriers, whereas flatter bands yield large *m\** and low mobility carriers.

The DOS measures the number of electronic states at a particular energy level that electrons are allowed to occupy, providing fundamental understanding of the material's electronic structure [72]. We can identify which atomic orbitals contribute to the valence and conduction bands and thus understand the origins of optical transitions and carrier behavior by examining the DOS. A high DOS at a band edge indicates a large number of available states for electrons or holes, which in turn influences conductivity and optical absorption.

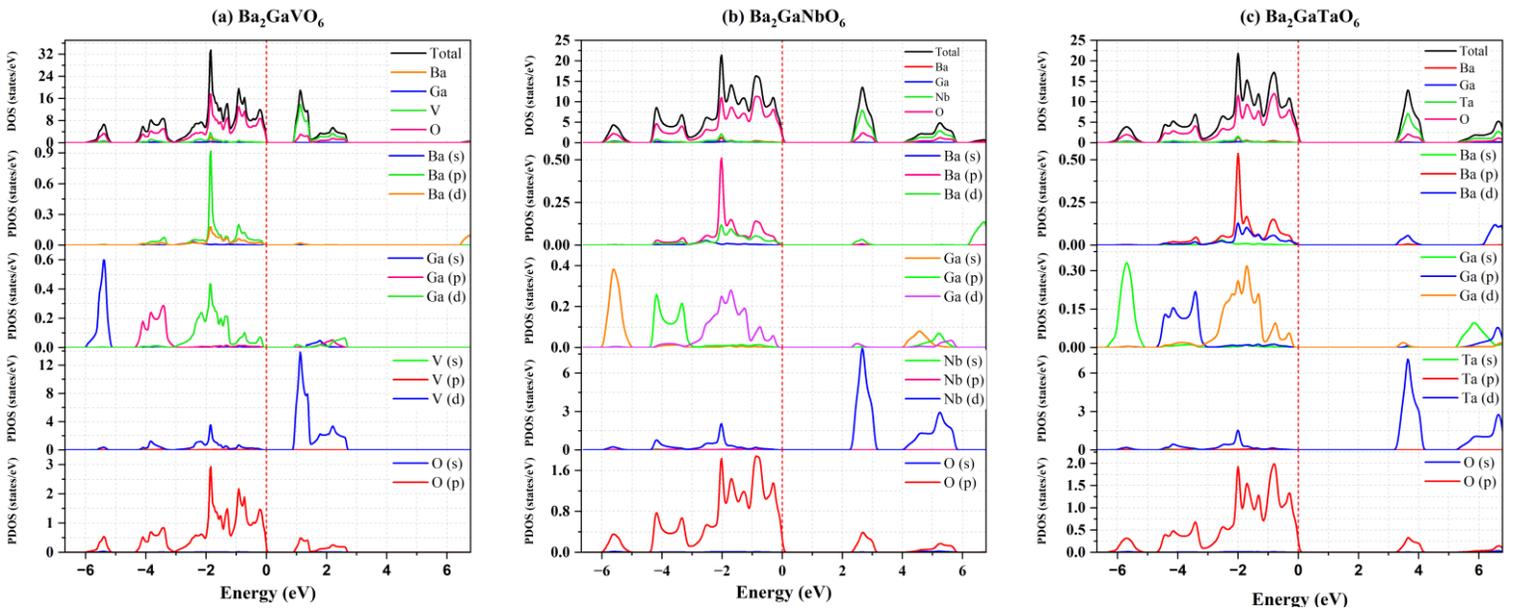

**Fig. 6.** TDOS and PDOS of (a) $Ba_2GaVO_6$, (b) $Ba_2GaNbO_6$, and (c) $Ba_2GaTaO_6$.

A detailed analysis of the total density of states (TDOS) and partial density of states (PDOS) provides deep insight into the electronic and optical properties of the material. The calculated TDOS and PDOS for $Ba_2GaXO_6$ (X = V, Nb, Ta) are shown in Fig. 6, spanning the energy range from −6.69 to 6.78 eV, using the TB-mBJ functional. All three compounds display multiple sharp peaks across the valence and conduction bands. The TDOS at the Fermi level ($E_F$) is essentially zero for all three compounds, confirming their semiconducting nature [57]. The gap separates the *O-2p*-derived valence band from the *X-d*-derived conduction band. With no states at $E_F$, the undoped material is insulating at low temperature and requires thermal or optical excitation to generate carriers. In practical terms, holes will populate the Oxygen valence states, and electrons will populate the *X*-cation (X = V, Nb, Ta) conduction states.

In the deep valence region (−6 to −2 eV), several intense features appear, while the TDOS rises in the conduction band above the Fermi level, reflecting the available conduction states. The valence band edge is predominantly dominated by O-*2p* orbitals. PDOS shows that *O−2p* states carry the bulk of the TDOS in the upper valence band. Because *O* is highly electronegative, *O−2p* orbitals lie highest in energy among the occupied states, forming filled bands at the top of the valence band. Conversely, *B*-site atom *Ga* contributes only a small amount of *Ga−4p* character at lower energy, and *A*-site atom *Ba* states lie at much deeper energies and do not affect the valence band maximum. Thus, any holes created by excitation will primarily reside on *O−2p* orbitals near the top of the valence band. The conduction band edge is mainly determined by the empty *d* orbitals of the *X*-site cations (X = V, Nb, Ta). The PDOS indicates that, just above the Fermi level, the TDOS is primarily contributed by X−d orbitals, which constitute the lower region of the conduction band. These metal *d* states also hybridize significantly with *O−2p* orbitals, as is typical for oxide perovskites [14]. Minor contributions from *Ga* or *Ba* orbitals occur only at higher energies well above the band edge. Consequently, electrons excited into the conduction band will occupy the *X−d* derived states, so that the primary optical transitions near the band gap are from *O−2p* to *X−d* (X = V, Nb, Ta).

For a more detailed insight into the bonding characteristics of the compound, the charge density mapping along the (100) crystallographic plane was examined employing the TB-mBJ functional for $Ba_2GaXO_6$ (X = V, Nb, Ta) compounds, as illustrated in Fig. 7. In all three compounds, the highest electron density occurs near the oxygen atoms and along the *Ga−O* and *X−O* (X = V, Nb,

Ta) bond directions, whereas the *Ba* sites appear as low-density regions. This distribution implies that *Ba−O* bonding is essentially ionic, while significant charge localization between *Ga/X* and *O* indicates partial covalent character. The overall charge distribution thus reflects a mixed ionic-covalent bonding nature [73].

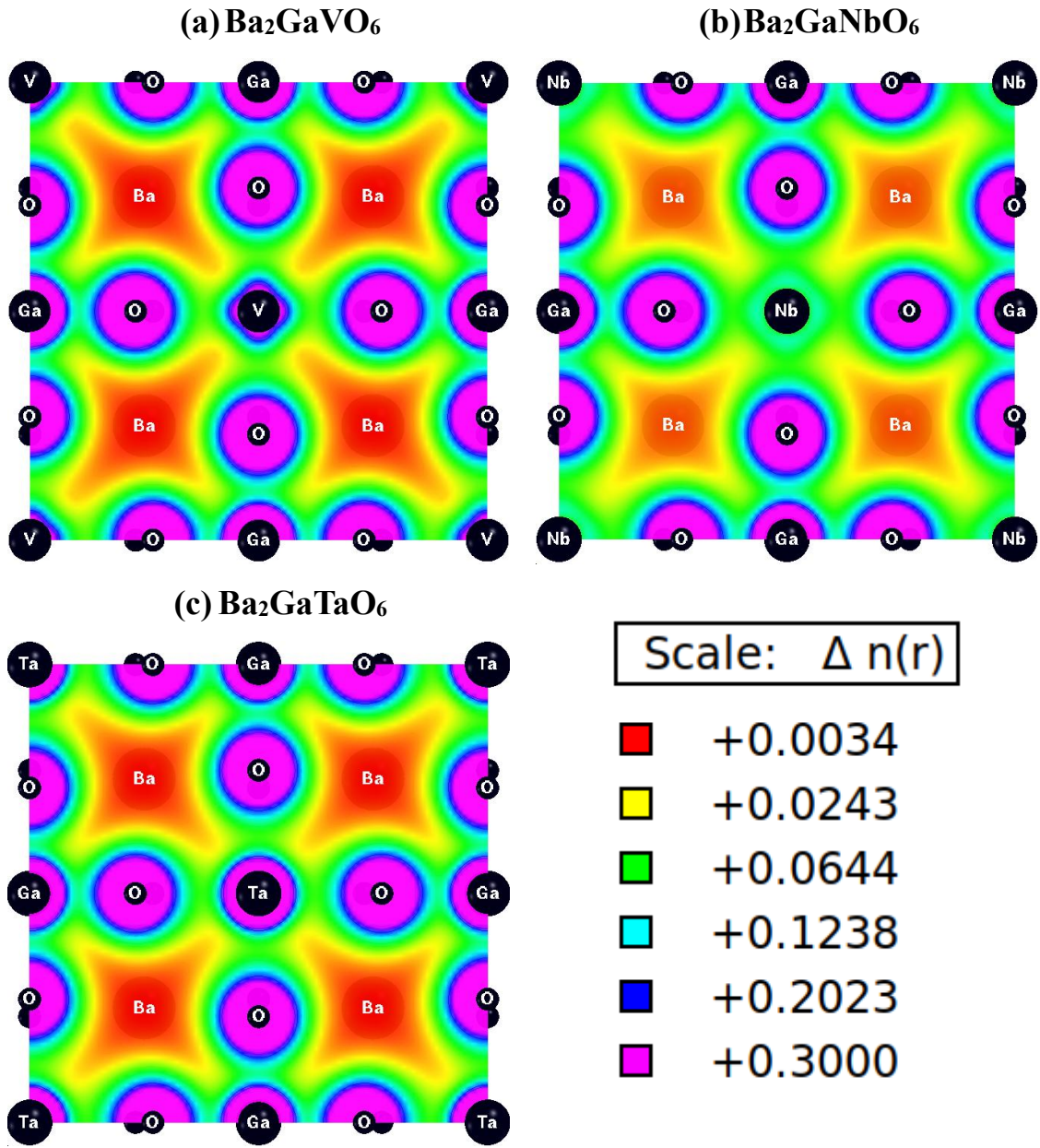

**Fig. 7.** Charge density mapping of (a) $Ba_2GaVO_6$, (b) $Ba_2GaNbO_6$ and (c) $Ba_2GaTaO_6$

The *Ga−O* bonds exhibit pronounced electron accumulation between *Ga* and *O*, consistent with *Ga*'s moderate electronegativity ($\chi \approx 1.81$) relative to oxygen ($\chi \approx 3.44$) and hence substantial covalent character. The *X−O* bonds (X = V, Nb, Ta) also show enhanced density near *O* but with comparatively smaller interatomic charge buildup, reflecting larger electronegativity differences. For *V* ($\chi \approx 1.63$) and *Nb* ($\chi \approx 1.60$), the $\Delta\chi$ with *O* is $\approx 1.8$, whereas for *Ta* ($\chi \approx 1.50$), $\Delta\chi$ is 1.9. This implies that *V−O* and *Nb−O* bonds retain slightly more covalency than *Ta−O*, as seen in Fig. 6, by a marginally higher density between *V/Nb* and *O* than for *Ta−O*. Ba$_2$GaXO$_6$ features predominantly ionic *Ba−O* bonds together with partially covalent *Ga−O* and *X−O* bonds. These observations are consistent with the electronegativity trends and previous reports of bonding in related perovskite oxides [74]. This trend correlates with the increasing band gap (0.92 → 2.3 → 3.23 eV). Smaller *B*-site cations generally give smaller gaps [75] because their *d* orbitals overlap more strongly with *O−2p*. Thus, Ba$_2$GaVO$_6$ ($V^{5+}$ is small) has the most delocalized charge distribution and narrowest gap, whereas Ba$_2$GaTaO$_6$ ($Ta^{5+}$ is larger) shows more localized density and a larger gap.

### 3.3 Thermoelectric transport properties:

Thermoelectric (*TE*) materials can convert waste heat into electrical energy, a key property that enhances their functionality and enables many practical applications across industries, thereby improving energy sustainability and addressing energy-related challenges. In semiconductors, the band structure determines transport properties, particularly the band gap relative to the Fermi level ($E_F$), carrier type and concentration, and the carrier effective masses, which contribute significantly [76]. To evaluate the capability of a compound for *TE* applications, it is necessary to investigate its transport properties using the Fermi energy determined through the TB−mBJ approach. In this study, transport coefficients were calculated using the BoltzTraP2 package, which employs a constant relaxation time approach with $\tau = 10^{-14}$ s, consistent with the standard thermoelectric calculation [77]. The temperature-dependent transport properties were evaluated over the range of 100−1800 K and are presented in Figs. 8 and 9.

Electrical conductivity ($\sigma$) represents the ability of a material to transport electric charge. In semiconductors, $\sigma$ depends on the charge carrier concentration (*N*) and mobility, and is expressed as: $\sigma = Ne\mu$ [78]. As the intrinsic carrier concentration, *N* grows exponentially as temperature increases according to $N \propto e^{\frac{-E_g}{2KT}}$. Thus, even a modest decrease in band gap produces a very large rise in $N$ at high temperature and increases the electrical conductivity. In our case, $Ba_2GaVO_6$ has the smallest gap ($E_g = 0.924$ eV) compared to $Ba_2GaNbO_6$ (2.354 eV) and $Ba_2GaTaO_6$ (3.279 eV). This means that at 1500 K, the intrinsic electron-hole concentration in $Ba_2GaVO_6$ is higher than that of the $Nb$ and $Ta$ analogues. The $Ba_2GaVO_6$ shows a higher $\sigma$ than the other two compounds. Our calculated values are consistent with the trend shown in Fig. 8(a). At 300 K temperature, reported values of $\sigma/\tau$ $(\Omega.m.s)^{-1}$ for $Ba_2GaVO_6$, $Ba_2GaNbO_6$, and $Ba_2GaTaO_6$ are $1.37\times10^{19}$, $1.17\times10^{19}$, and $1.29\times10^{19}$ $(\Omega.m.s)^{-1}$. Thus, at room temperature, $Ba_2GaNbO_6$ shows the highest conductivity, closely followed by $Ba_2GaVO_6$, while $Ba_2GaTaO_6$ is slightly lower. At 1500 K, $Ba_2GaVO_6$ rises to ($\sigma/\tau \approx 5.53\times10^{19}$) and exceeds $Ba_2GaNbO_6$ ($\sigma/\tau \approx 3.36\times10^{19}$) and $Ba_2GaTaO_6$ ($\sigma/\tau \approx 3.65\times10^{19}$). The monotonic increase of $\sigma$ with temperature in all three materials is consistent with $\sigma \propto N(e\mu)$ [79]. In addition, the carrier effective mass controls mobility. Comparing the three oxides: $Ba_2GaVO_6$ has a moderate electron mass ($m_e^* = 0.79$) but very light holes ($m_h^* = 0.25$). Thus, $Ba_2GaVO_6$ benefits from its huge $N$ and good hole mobility; its heavier electrons are more than offset by the extremely light holes. By contrast, $Ba_2GaTaO_6$, despite its light electrons, has a much smaller $N$, due to $E_g \sim 3.28$ eV and very heavy holes, so its overall $\sigma$ is lowest. These high $\sigma$ values ($\sim 10^{19}$ $(\Omega.m.s)^{-1}$ at room temperature and even higher at 1500 K) and strong temperature dependence suggest excellent charge transport. Such behavior is promising for thermoelectric and waste heat applications, where efficient charge and heat flow are needed [78]. Our computed results are consistent with trends reported for other high $ZT$ materials. Notably, the electrical conductivity of $Ca_2ZrTiO_6$ increases with temperature, reaching $\sigma = 6.2\times10^4$ $(\Omega.m)^{-1}$ at 1200 K [1]. $BaIn_2C_2$ and $SrIn_2C_2$ exhibit high electrical conductivity, reported as $0.135\times10^{20}$ $(\Omega.m.s)^{-1}$ and $0.15\times10^{20}$ $(\Omega.m.s)^{-1}$, respectively [68]. These high conductivity values, together with the calculated Seebeck coefficients and thermal conductivities, contribute to the elevated predicted $ZT$. Higher electrical conductivity ($\sigma$) improves the power factor ($S^2.\sigma$) and boosts electronic thermal conductivity, which supports the Wiedemann−Franz law: $k_e = LT\sigma$, where $L$ is the Lorentz constant ($2.44\times10^{-8}$ $W/S.K^2$) [79]. $Ba_2GaXO_6$ (X = V, Nb, Ta) materials show significantly large $\sigma$ with temperature, which is beneficial for converting heat to electricity [56].

As depicted in Fig. 8(b), the electronic thermal conductivity ($k_e/\tau$) rises with temperature for all three compounds. At room temperature (300 K), $k_e/\tau$ for $Ba_2GaVO_6$, $Ba_2GaNbO_6$, and $Ba_2GaTaO_6$ are roughly $0.079\times10^{15}$, $0.055\times10^{15}$ and $0.061\times10^{15}$ $W.m^{-1}.K^{-1}.s^{-1}$, respectively. All three values rise to about $1.05\times10^{15}$, $0.33\times10^{15}$, and $0.39\times10^{15}$ $W.m^{-1}.K^{-1}.s^{-1}$ by 1500 K. $Ba_2GaNbO_6$ exhibits an ultralow electronic thermal conductivity ($k_e/\tau$) at elevated temperatures compared to other compounds. This steady increase follows the Wiedemann-Franz law ($k_e \propto \sigma T$) [80], since higher temperature thermally excites more charge carriers to carry heat.

Thermal conductivity, representing a material's ability to conduct heat, comprises both electronic and lattice (phonon) contributions [81]. The phonon dispersion relations of $Ba_2GaXO_6$ (X = V, Nb, Ta), shown in Fig. 3, reveal characteristic differences that impact lattice thermal conductivity. In general, thermal transport is governed by the group velocities of acoustic phonons and by anharmonic scattering among phonons. High elastic stiffness and phonon frequencies yield large group velocities and high thermal conductivity.

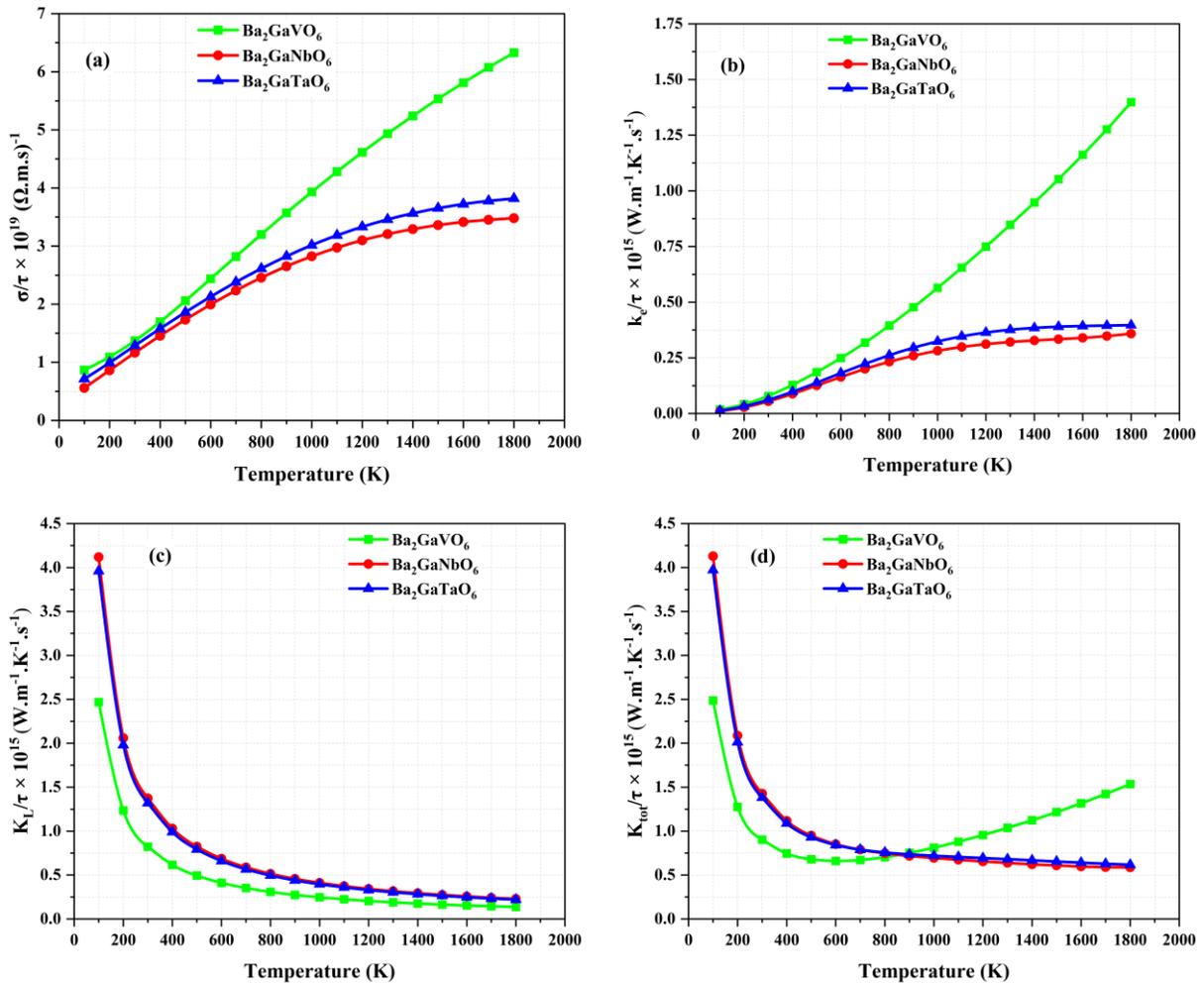

**Fig. 8.** The calculated (a) $\sigma/\tau$, Electrical conductivity, (b) $k_e/\tau$, Electronic thermal conductivity, (c) $k_L/\tau$, Lattice thermal conductivity, (d) $k_{tot}/\tau$, Total thermal conductivity of the studied compounds.

When low-frequency optical modes overlap strongly with acoustic modes, the increased acoustic-optic scattering greatly suppresses thermal conductivity. Total thermal conductivity ($k_{tot}$) combines both the lattice thermal conductivity ($k_L$) and the electronic thermal conductivity ($k_e$), expressed as: $k_{tot} = k_e + k_L$. BoltzTraP2 code addresses only the electronic component ($k_e/\tau$), since phonon-driven heat transport lies outside its intended computational scope. Therefore, Slack's model is commonly employed to independently estimate $k_L$, using the following relation [82]:

$$K_L = \frac{A(\gamma)\,\delta\,M_{av}\,\theta_D^3}{\gamma^2\,n^{\frac{2}{3}}\,T} \tag{8}$$

$\theta_D$ is the Debye temperature, $M_{av}$ denotes the average atomic mass, $A(\gamma)$ is a constant set by the Gruneisen parameter $\gamma$, $\delta$ equals the cubic root of the mean atomic volume, $T$ is the absolute temperature in Kelvin, $n$ is the number of atoms in the primitive unit cell, and $\upsilon$ is Poisson's ratio. In calculating the lattice thermal conductivity, $k_L$, it is crucial to consider the number of atoms, $n$, per primitive unit cell rather than per conventional unit cell. This is because Slack's model is derived on the basis of the primitive cell, where the atom count directly reflects the fundamental periodicity and vibrational degrees of freedom of the crystal lattice. Using the atom per conventional cell may yield an artificially higher $ZT$ value. Therefore, in accordance with Slack's equation, using the number of atoms per primitive unit cell is the correct approach. Here, $\gamma$ and $A(\gamma)$ can be determined from the following equations:

$$\gamma = \frac{3(1+\upsilon)}{2(2-3\upsilon)} \tag{9}$$

$$A(\gamma) = \frac{4.85628 \times 10^7}{2\left(1 - \frac{0.514}{\gamma} + \frac{0.228}{\gamma^2}\right)} \tag{10}$$

Fig. 8(c) presents the calculated lattice thermal conductivity, $k_L/\tau$. At 300 K temperature the value of $k_L/\tau$ is $0.82\times10^{15}$, $1.37\times10^{15}$, and $1.32\times10^{15}$ W.m$^{-1}$.K$^{-1}$.s$^{-1}$ for Ba$_2$GaVO$_6$, Ba$_2$GaNbO$_6$, and Ba$_2$GaTaO$_6$, respectively. By 1500 K, it drops to about $0.16\times10^{15}$, $0.27\times10^{15}$, and $0.26\times10^{15}$ W.m$^{-1}$.K$^{-1}$.s$^{-1}$. $k_L/\tau$ decreases by more than an order of magnitude over 100 − 1100 K. At temperatures above 1100 K, the lattice thermal conductivity ($k_L$) decreases and approaches a constant value. This data follows the inverse relationship between temperature and $k_L$, consistent with Slack's model ($k_L \propto 1/T$) [83].

Fig. 8(d) illustrates the total thermal conductivity, $k_{tot}/\tau$. At a temperature of 300 K, the value of $k_{tot}/\tau$ is $0.90 \times 10^{15}$, $1.43 \times 10^{15}$, $1.38 \times 10^{15}$ W.m$^{-1}$.K$^{-1}$.s$^{-1}$ for the Ba$_2$GaVO$_6$, Ba$_2$GaNbO$_6$, and Ba$_2$GaTaO$_6$ compounds (almost equal to their $k_L$ values). As the temperature rises, $k_{tot}$ first drops with the falling lattice thermal conductivity. Above ~600 K, $k_{tot}/\tau$ climbs again for Ba$_2$GaVO$_6$ because the electronic term ($k_e \propto \sigma T$) grows and begins to dominate [80]. This crossover behavior is expected according to the Wiedemann-Franz relation. However, for Ba$_2$GaNbO$_6$ and Ba$_2$GaTaO$_6$, $k_{tot}/\tau$ remains stable above 600 K because the electronic thermal conductivity is very low at high temperatures for these compounds. The moderate values of $k_{tot}/\tau$ imply that these oxides are relatively poor heat conductors, which helps maintain a thermal gradient in a thermoelectric device.

The Seebeck coefficient (S) measures the voltage generated by a temperature gradient and reveals the dominant charge carriers [84]. A positive $S$ indicates holes (p-type) dominate, while a negative $S$ indicates electrons (n-type) dominate the transport properties [56]. For Ba$_2$GaXO$_6$ (X = V, Nb, Ta), $S$ is positive over the entire 100 −1800 K range for all compounds, confirming p-type behavior, depicted in Fig. 9(a). At 300 K temperature, reported Seebeck coefficient values for Ba$_2$GaVO$_6$, Ba$_2$GaNbO$_6$, and Ba$_2$GaTaO$_6$ are 171.2, 184.3, and 176.8 $\mu$V/K. Ba$_2$GaVO$_6$ consistently shows the largest $S$, while Ba$_2$GaTaO$_6$ is the lowest; this ordering persists at other temperatures. All Seebeck coefficient values decrease with rising temperature because increasing carrier concentration and thermal excitation reduce the thermoelectric voltage per degree [85]. The Seebeck coefficient and charge carrier concentration are inversely correlated, as expressed by the following equation [86]:

$$S = \left(\frac{8\pi^2 k_B^2}{3h^2 e}\right) \left(\frac{\pi}{3N}\right)^{\frac{2}{3}} m^* T \qquad (11)$$

Here $m^*$ is the effective mass. The Seebeck coefficient, S, is strongly influenced by the carrier effective mass. $S$ increases with the carrier effective mass because a heavier mass raises the density of states near the Fermi level ($E_F$). A large band effective mass tends to enhance the thermopower, whereas a light mass leads to a lower S. This trend is evident in our results, as illustrated in Table 2. Ba$_2$GaVO$_6$ has the highest effective mass and accordingly exhibits the largest Seebeck coefficient, while Ba$_2$GaNbO$_6$ and Ba$_2$GaTaO$_6$ have a low Seebeck coefficient value. These observations are consistent with the theory presented in equation (11). The high Seebeck coefficient of Ba$_2$GaVO$_6$ is a result of its high effective mass, which aligns with the expected direct correlation between heavy band mass and large *S* in thermoelectric materials. Seebeck coefficient values greater than 250 µV/K are rare; however, values above ~200 µV/K are considered suitable for thermoelectrics [87]. Hadji et al. [88] reported that $Cs_2CaGeI_6$ attains a high thermoelectric figure of merit, $ZT = 2.5$ at 1000 K, with a calculated Seebeck coefficient in the range 100–174 µV/K over 200−1000 K. $SrIn_2C_2$ was reported to exhibit Seebeck coefficients of 128−206 µV/K between 300−*1000 K*, with a peak *ZT* of 1.93 [68]. $BaIn_2C_2$ exhibits particularly large Seebeck values of 225−244.45 µV/K in the 300−1000 K interval and a reported maximum ZT of approximately 2.86 [68]. An extremely high Seebeck coefficient tends to coincide with low $\sigma$, so a balance is needed. Our materials strike a reasonable compromise, as all exhibit S on the order of 150−210 µV/K with high conductivity, yielding promising power factors. This makes them strong candidates for waste heat recovery and thermoelectric generators.

The thermoelectric power factor (*PF*) quantifies a material's ability to generate electrical power efficiently from thermal energy. It is defined as $PF = S^2 \cdot \sigma/\tau$ [89]. For efficient thermoelectric power generation, a high-power factor is essential [89]. Fig. 9(b) shows the calculated PF value. The *PF* rises monotonically with temperature for all three compounds, a common behavior as thermal excitation increases carrier concentration. At 300 K, the reported value of *PF* is $0.402 \times 10^{12}$, $0.396 \times 10^{12}$, and $0.403 \times 10^{12}$ W.m$^{-1}$.K$^{-2}$.s$^{-1}$ for $Ba_2GaVO_6$, $Ba_2GaNbO_6$, and $Ba_2GaTaO_6$, respectively, increasing to about $2 \times 10^{12}$, $0.65 \times 10^{12}$, and $0.77 \times 10^{12}$ W.m$^{-1}$.K$^{-2}$.s$^{-1}$ by 1700 K. All materials show a temperature-dependent gain reflecting enhanced carrier transport at high temperature. These large power factor values indicate that $Ba_2GaXO_6$ (X = V, Nb, Ta) compounds have strong potential for thermoelectric power generation [56].

**Table 3.** Thermoelectric (*TE*) transport properties of $Ba_2GaXO_6$ (X = V, Nb, Ta) at 300 K and ZT values at different temperatures:

| Compound | $\sigma/\tau \times 10^{19}$ ($\Omega$.m.s)$^{-1}$ | $k_e/\tau \times 10^{15}$ (W.m$^{-1}$.K$^{-1}$.s$^{-1}$) | $k_L/\tau \times 10^{15}$ (W.m$^{-1}$.K$^{-1}$.s$^{-1}$) | S (µV/K) | PF×10$^{12}$ (W.m$^{-1}$.K$^{-2}$.s$^{-1}$) | ZT value at 600 K | 1500 K | 1800 K |
|---|---|---|---|---|---|---|---|---|
| $Ba_2GaVO_6$ | 1.37 | 0.079 | 0.82 | 171.2 | 0.402 | 0.86 | 2.36 | 2.38 |
| $Ba_2GaNbO_6$ | 1.17 | 0.055 | 1.37 | 184.3 | 0.396 | 0.48 | 1.78 | 1.88 |
| $Ba_2GaTaO_6$ | 1.29 | 0.061 | 1.32 | 176.8 | 0.403 | 0.50 | 1.91 | 2.14 |

The figure of merit, ZT, is a dimensionless parameter that quantifies how effectively a thermoelectric material converts heat into electrical energy, defined by, $ZT = \frac{S^2 \sigma T}{k_e + k_L}$, where ($k_e + k_L$) is the total thermal conductivity [90]. Materials exhibiting higher ZT values (above 1) offer superior performance in thermoelectric applications, such as waste heat recovery generators or cooling systems [91]. Fig. 9(c) presents the calculated ZT values. At a temperature of 300 K, the ZT values are relatively low, 0.13, 0.08, and 0.09 for Ba2GaVO6, Ba2GaNbO6, and Ba2GaTaO6, respectively. However, above 300 K, the ZT values climb rapidly. For $Ba_2GaVO_6$, ZT surpasses 1.0 at ~ 700 K (ZT ≈ 1.16) and reaches ~2.36 by 1500 K. The $Ba_2GaNbO_6$ and $Ba_2GaTaO_6$ compounds cross ZT = 1 at 900 K, respectively, and achieve 1.78 and 1.91 by 1500 K, respectively. As the temperature approaches the compounds' melting point (≈ 1800 K), the thermoelectric figure of merit (ZT) begins to decrease, which may be attributable to enhanced bipolar conduction and increased phonon-phonon scattering at very high temperatures.

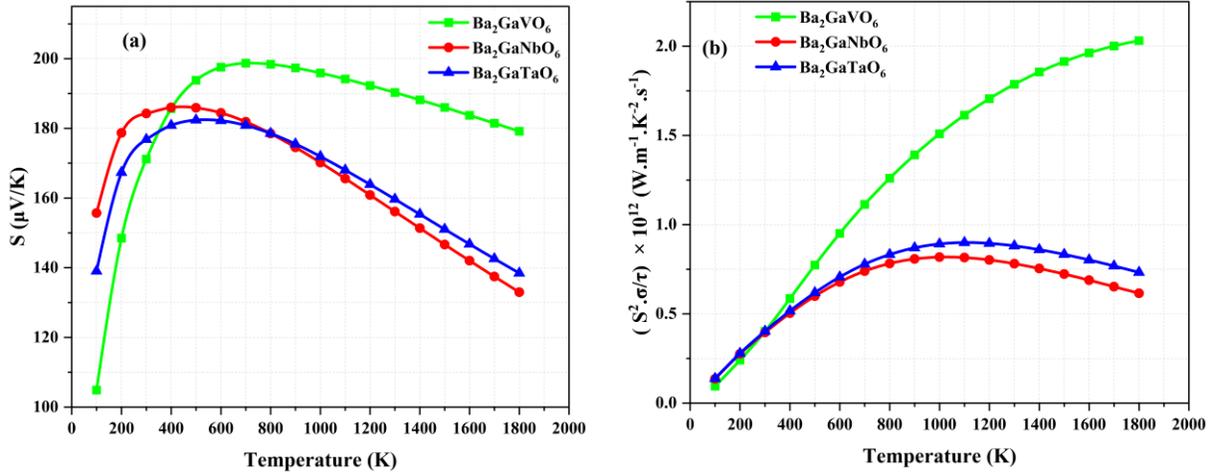

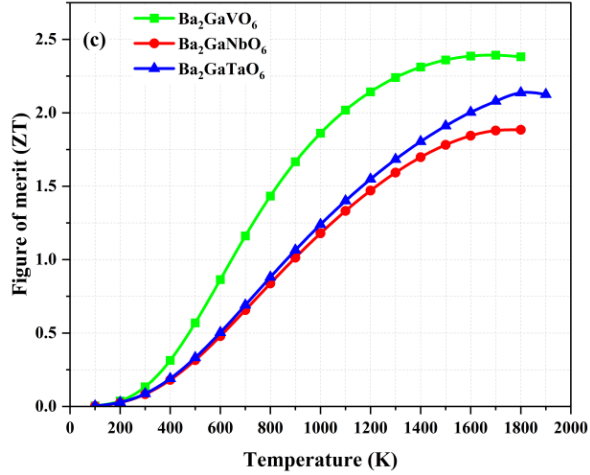

**Fig. 9.** The computed (a) *S*, Seebeck coefficient, (b) Power factor, (c) *ZT*, Figure of merit of the studied compounds.

$Ba_2GaXO_6$ (X = V, Nb, Ta) displays a combination of transport properties that support the measured *ZT* values above 1. The compounds exhibit a large Seebeck coefficient, together with very high electrical conductivity, on the order of $10^{19}$, shown in Fig. 8(a) and Fig. 9(a), producing a substantial power factor $S^2\sigma$. Simultaneously, the ultra-low thermal conductivity ($k_{tot}$), because of phonon scattering at high temperatures, is the primary reason for the high figure of merit (*ZT*) value. At high temperatures (above 1000 K), $Ba_2GaVO_6$ exhibits a higher *ZT* value compared to $Ba_2GaNbO_6$ and $Ba_2GaTaO_6$, due to its superior *PF* and smaller band gap. This enhancement primarily arises from its Seebeck coefficient, which is approximately 14% higher than $Ba_2GaNbO_6$ and $Ba_2GaTaO_6$ in this temperature range. $Ba_2GaVO_6$ demonstrates significantly lower thermal conductivity at high temperatures, a critical factor for achieving superior thermoelectric performance. All three compounds show monotonic *ZT* growth with temperature. Thermoelectric materials, such as $Bi_2Te_3$ or SiGe, typically have a *ZT* ≈ 1 at their optimum temperatures [92]. Our calculated values are comparable with other compounds such as $Cr_2ZrTiO_6$ (ZT = 4.4) [1], $Ba_2FeMoO_6$ (*ZT* = 0.998) [28], $Sr_2HoNbO_6$ (*ZT* = 0.97) [20], $Cs_2CaGeI_6$ (*ZT* = 2.5) [88], $MgAl_2C_2$ (ZT = 1.2) [68], $SrIn_2C_2$ (ZT = 1.93) [68], and $BaIn_2C_2$ (ZT = 2.86) [68]. The *ZT* values above 2 predicted here are thus very significant and indicate excellent high-temperature thermoelectric performance, implying that as temperature increases, these materials become increasingly efficient in converting heat to electricity. It should be noted that Slack's equation typically provides lattice thermal conductivity ($k_L$) values higher than the actual ones [91], except for some cases, such as

SnSe [89] and Ag$_2$XYSe$_4$ [93], where the predictions for $k_\text{L}$ closely agree with experimental measurements. Thus, in the present case, one can expect experimentally measured ZT values to exceed those calculated here.

### 3.4 Mechanical properties:

All calculated elastic constants of Ba$_2$GaVO$_6$, Ba$_2$GaNbO$_6$, and Ba$_2$GaTaO$_6$ meet the Born mechanical stability conditions for cubic crystals [94]:

$C_{11} > 0$, $C_{12} > 0$, $C_{44} > 0$, $C_{12} < B < C_{11}$, $C_{11} - C_{12} > 0$, and $C_{11} + 2C_{12} > 0$.

In each case, the computed values meet the mechanical stability conditions, as shown in Table 4, indicating that all three compounds are mechanically stable in their predicted structures. In every case, $C_{11}$ is significantly larger than both $C_{12}$ and $C_{44}$, which implies that the crystal is far stiffer against uniaxial compression along the principal axes than against shear deformation. This pattern means that each material will resist volume changes under pressure more strongly than it will resist shape changes under shear [95]. Ba$_2$GaTaO$_6$ exhibits the largest $C_{11}$ and $C_{44}$; its high $C_{11}$ and $C_{44}$ values indicate a very strong resistance to both axial compression and shear. The difference ($C_{11} - C_{12}$) is much smaller for Ba$_2$GaVO$_6$ (≈ 75.8 GPa) than for the Ba$_2$GaNbO$_6$ and Ba$_2$GaTaO$_6$ analogues, suggesting that Ba$_2$GaVO$_6$ is relatively more compressible under axial strain. Overall, Ba$_2$GaTaO$_6$ should exhibit the greatest resistance to uniaxial and shear deformations, followed by Ba$_2$GaNbO$_6$ and then Ba$_2$GaVO$_6$, which is slightly less stiff by comparison.

**Table 4:** Calculated elastic parameters of Ba$_2$GaXO$_6$ (X = V, Nb, Ta) compounds:

| Parameter | Ba$_2$GaVO$_6$ | Ba$_2$GaNbO$_6$ | Ba$_2$GaTaO$_6$ |
|---|---|---|---|
| $C_{11}$ | 214.695 | 226.976 | 235.581 |
| $C_{12}$ | 138.866 | 113.775 | 120.61 |
| $C_{44}$ | 108.036 | 108.116 | 112.814 |
| Bulk Modulus, $B$ (GPa) | 164.142 | 151.509 | 158.934 |
| Poisson's Ratio, $\upsilon$ | 0.311 | 0.27 | 0.271 |
| Entropy, $S$ (J/K.mol) | 233.84 | 221.54 | 225.13 |
| Debye temperature, $\Theta_D$ (K) | 477.5 | 496 | 466.6 |
| Melting temperature, $T_m$ (K) | 1822 | 1894 | 1945 |

## 4. Conclusions

This study successfully predicts promising thermoelectric double perovskite oxides, $Ba_2GaXO_6$ (X = V, Nb, Ta), using first-principles calculations, demonstrating that they are stable and can be synthesized. Electronic structure calculations employing the TB−mBJ potential reveal direct band gaps of 0.924 eV ($Ba_2GaVO_6$), 2.354 eV ($Ba_2GaNbO_6$), and 3.279 eV ($Ba_2GaTaO_6$), suggesting these compounds are promising semiconductor materials. Mechanical stability is confirmed by the calculated elastic constants ($C_{11}$, $C_{12}$, and $C_{44}$), which also indicate ductile behavior and ionic bonding, with Debye temperatures consistent with strong lattice frameworks. The calculated transport properties from the BoltzTrap2 code indicate high electrical conductivity — approximately $10^{19}$ S/m —which increases significantly with temperature across all three compounds. A relatively large Seebeck coefficient was observed, with *S* positive throughout the 100−1800 K range for all compounds, confirming their *p*-type behavior. The combination of high electrical conductivity and Seebeck coefficient, along with ultralow lattice thermal conductivity at high temperatures, leads to very high *ZT* values of 2.36, 1.78, and 1.91 at 1500 K for $Ba_2GaVO_6$, $Ba_2GaNbO_6$, and $Ba_2GaTaO_6$, respectively, highlighting their excellent potential for converting waste heat into electricity. It should be noted that the predicted compounds show moderate *ZT* values of 0.86, 0.48, and 0.50 at 600 K, and very high *ZT* values of 2.38, 1.88, and 2.14 at 1800 K [Close to the melting point; Table 4] for $Ba_2GaVO_6$, $Ba_2GaNbO_6$, and $Ba_2GaTaO_6$, respectively. Therefore, these materials are expected to be synthesized and hold great promise as environmentally friendly options for future renewable energy uses, such as thermoelectric devices.

**Acknowledgement:**

This work was carried out at the ACMRL, which was established with a research grant (grant number: 21-378 RG/PHYS/AS_G-FR3240319526) from UNESCO-TWAS and the Swedish International Development Cooperation Agency (SIDA).

**CRediT Author contributions**

**S. S. Saif:** Data curation, Investigation, Visualization, Formal analysis, Writing – original draft.
**M. M. Hossain:** Formal analysis, Validation, Writing - review & editing. **M. A. Ali:** Conceptualization, Methodology, Formal analysis, Validation, Project administration, Writing – review & editing, Supervision.